\documentclass[english,english,noshowpacs,amssymb,aps,nofootinbib]{revtex4-2}

\usepackage{amsmath}
\usepackage{graphicx}

\newcommand{\ket}[1]{|{#1}\rangle} \newcommand{\bra}[1]{\langle{#1}|}

\begin{document}

\title{Deterministic Generation of Genuine Tri-Partite\\ Hybrid Atom-Photon Entanglement through Dissipation}

\author{Pablo Barberis-Blostein}
\affiliation{Instituto de Investigaciones en Matem\'aticas Aplicadas y en Sistemas. Universidad
  Nacional Aut\'onoma de M\'exico, Ciudad Universitaria, 04510 M\'exico D.F. M\'exico}
\author{Alberto M. Marino}
\affiliation{Homer L. Dodge Department of Physics and Astronomy, The University of Oklahoma, Norman, Oklahoma 73019, USA}
\affiliation{Center for Quantum Research and Technology, The University of Oklahoma, Norman, Oklahoma 73019, USA}

\begin{abstract}
The ability to deterministically generate genuine multi-partite entanglement is fundamental for the advancement of quantum information science. We show that the interaction between entangled twin beams of light and an atomic ensemble under conditions for electromagnetically induced transparency leads to the generation of genuine hybrid tri-partite entanglement between the two input fields and the atomic ensemble. In such a configuration, the system is driven through dissipation to a steady state given by the hybrid entangled state. To show the presence of the genuine hybrid entanglement, we introduce a new approach to treat the atomic operators that makes it possible to show a violation of a tri-partite entanglement criterion based on the properties of the two optical fields and collective properties of the atomic ensemble. Additionally, we show that while each of the input optical fields does not exhibit single beam quadrature squeezing, as the fields propagate through the atomic medium their individual quadratures can become squeezed and in some cases oscillate between the presence and absence of squeezing.  Finally, we propose a technique to characterize the tri-partite entanglement through joint measurements of the fields leaving the atomic medium, making such an approach experimentally accessible.
\end{abstract}

\maketitle

\section{Introduction}

Entanglement is at the heart of the emerging field of quantum technologies~\cite{rsta.2003.1227}.  As such, it is necessary to find new and robust ways to deterministically generate it. In many applications in quantum information science that range for quantum communications~\cite{PhysRevX.9.041042} to quantum computing~\cite{RN1173,9149990} to distributed quantum sensing~\cite{acsphotonics.9b00250,NatPhys.16.281,Zhuang_2020}, entanglement needs to be distributed between a network of quantum systems.  For the implementation of such quantum networks the generation of entangled states of light with atomic systems~\cite{McCormick:08,Boyer_2008,Ding:15,Wang:17,Park:19,Wang:20} and hybrid light-atom entanglement~\cite{PhysRevLett.124.010510} offer a promising approach given that to date the most robust and efficient memories~\cite{RN1172,09500340.2016.1148212} and repeaters~\cite{RevModPhys.83.33} are based on atomic systems.  Furthermore, the extension to multi-partite entanglement can enable or enhance applications such as quantum sensor networks~\cite{Ren:12,Eldredge:18} and multi-party quantum communications~\cite{Hillery:99,Zhu:15}.

We propose a novel mechanism based on electromagnetically induced transparency (EIT) for the generation of genuine tri-partite hybrid entanglement between two optical fields and an atomic ensemble. It is well known that in EIT the interaction between two optical fields and an atomic system in a $\Lambda$ configuration drives the atom-field system into a dark state in which the fields can propagate through the medium without absorption~\cite{RevModPhys.77.633,rv:marangos}.  Once such a dark state is established the system reaches steady state and the mean values of the fields propagating in an ideal EIT medium do not change; however, their statistical properties do. It has been previously shown that phase fluctuations in each of the two fields that enter the EIT medium become correlated as the fields propagate~\cite{rv:richter}. These results were later extended to the quantum domain~\cite{rv:pablomarc2} and it was shown that for the case of separable quantum states of the light propagating through an EIT medium the statistical properties of the optical fields are modified with propagation, but no entanglement is generated.

To the best of our knowledge, there have been no studies on the evolution of entangled fields in an EIT medium. To tackle this problem, we consider the interaction between two entangled optical fields, in the form of twin beams, with an atomic ensemble in a $\Lambda$ configuration. We show that EIT provides a mechanism for the redistribution of the initial entanglement in the fields to a hybrid atom-field entanglement through dissipation. As the light propagates through the atomic medium,  due to dissipation introduced by atomic spontaneous emission, the system reaches a stationary state given by a tri-partite entangled state between the two input optical fields and the atomic ensemble. This process leads to the deterministic and unconditional generation of the entangled state. Given that the generation of entanglement through dissipation does not require the preparation of the atomic system in a particular state, it makes this technique inherently stable against perturbations~\cite{PhysRevA.81.043802,PhysRevLett.107.080503,PhysRevA.83.052312,Stannigel.2012}. As a result, this new entanglement mechanism is robust and can be implemented with current technology, as it only requires the propagation of the input entangled fields through the atomic medium.

To verify the presence of genuine tri-partite entanglement we introduce a new approach to treat the atomic operators that allows us to establish a set of inequalities based on the properties of the optical fields and collective atomic variables. Furthermore, we identify a measurement protocol to verify these inequality that is experimentally accessible and only requires measurements of the individual and collective noise properties of the output optical fields.

\section{Theoretical description of the system}
\label{sc:equations}

\begin{figure}[h]
\centering
  \includegraphics[width=5cm]{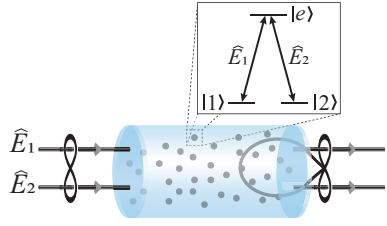}
  \caption{Schematic of the interaction between entangled twin beam of light and an atomic ensemble in a $\Lambda$ configuration in the EIT regime. Each of the entangled fields is assumed to be resonant with a single transition of the $\Lambda$ system.}
  \label{fig:setup}
\end{figure}

We consider an atomic ensemble in a $\Lambda$ configuration consisting of two ground states,
$|1\rangle$ and $|2\rangle$, dipolarly coupled to a common excited
state $|e\rangle$. We assume the atomic transitions from the ground states
$|i\rangle$ , $i=1,2$, to $|e\rangle$ to have the same frequency, $\omega_a$, and the transition between the two ground states to be dipole forbidden. Each of the atomic transitions interacts with a single resonant electromagnetic field, $\hat{\vec E}_j$, as shown in Fig.~\ref{fig:setup}. These fields can correspond to the two different polarization of the field, for example. We refer to $E_{1}$ as the control field and $E_{2}$ as the probe field. We write
the field operator as
\begin{equation}
\hat{\vec E}_j=\vec{\mathcal
  E}_j \hat a_j(z,t) \exp[ik_{{\rm c},j}z-\omega_{{\rm c},j} t] + {\rm h.c.}\, ,
\end{equation}
where
$|\vec{\mathcal E}_j|=\sqrt{\hbar\omega_{{\rm c},j}/(2\epsilon_0 V)}$
is the vacuum electric field at the carrier frequency
$\omega_{{\rm c},j}=c k_{{\rm c},j}$ in a medium with volume $V$, and
$\hat a_j(z,t)$ is the field envelope operator of the corresponding field
$j$.

To study the light-matter interaction under the configuration shown in Fig.~\ref{fig:setup}, we take the atomic medium to be one dimensional with length $L$ and consider the limit in which the inter-atomic distance is much smaller than the shortest relevant wavelength of the fields. In this limit, the atomic operators can be considered as continuous operators that are a function of  distance $z$ along the propagation direction of the fields such that
\begin{equation}\label{eq:oz}
\hat{\sigma}_{\mu\nu}(z) = \lim_{\Delta z\rightarrow 0} \frac{L}{N\Delta
              z}\!\sum\limits_{z_{j}\in \Delta z }\! \hat{\sigma}_{\mu\nu}^{(j)}\, ,
\end{equation}
where $\hat\sigma^{(l)}_{\mu\nu}=\ket{\mu}^{(l)}{}^{(l)}\bra{\nu}$ represents the transition operator of atom $l$ at position $z_l$ and $\Delta z$ is a small volume around position $z$.

The dynamics of the fields as they propagate through the atomic medium are governed by the driven wave equations, which can then be written as
\begin{equation}
\left(\frac{\partial}{\partial t}+c\frac{\partial}{\partial z}\right)\hat a_j = -i
g_j N \hat\sigma_{je}(z)\, ,
\label{eq:eprop}
\end{equation}
with $j=1,2$ and where we have introduced the coupling constant
$g_j=\vec \wp\cdot\vec{\mathcal E}_j/\hbar$ with atomic dipole moment $\vec \wp$. The dynamics of the atoms, on the other hand, are given by the
Heisenberg-Langevin equations
\begin{alignat}{1}
\frac{\partial}{\partial t} \hat\sigma_{z1} =&\frac{1}{3}(-\gamma_1-\gamma)(1+ \hat\sigma_{z1}+\hat\sigma_{z2})-2 i (\hat\sigma_{e1}g_1\hat a_1  - g_1^*\hat a^\dagger_1\hat\sigma_{1e}) - i (\hat\sigma_{e2}g_2\hat a_2-g_2^*\hat a^\dagger_2\hat\sigma_{2e})  + \hat f_{z1},\nonumber\\
\frac{\partial}{\partial t}\hat\sigma_{z2} =&\frac{1}{3}(-\gamma_2- \gamma)(1+\hat\sigma_{z1}+\hat\sigma_{z2})
-i (\hat\sigma_{e1} g_1\hat a_1  -g_1^*\hat a^\dagger_1
\hat\sigma_{1e}) - 2 i  (\hat\sigma_{e2} g_2\hat
a_2-g_2^*\hat a^\dagger_2 \hat\sigma_{2e}) +\hat f_{z2},\nonumber\\
\frac{\partial}{\partial t} \hat\sigma_{1e}=&-\frac{\gamma}{2}
\hat\sigma_{1e}+i g_1 \hat\sigma_{z1} \hat a_1-i g_2 \hat\sigma_{12} \hat a_2
+\hat f_{1e},\nonumber\\
\frac{\partial}{\partial t} \hat\sigma_{2e}=&-\frac{\gamma}{2}
\hat\sigma_{2e}+i g_2 \hat\sigma_{z2} \hat a_2   -i g_1 \hat\sigma_{21} \hat a_1
+ \hat f_{2e} ,\nonumber\\
\frac{\partial}{\partial t}\hat\sigma_{21}=&-i g_1^* \hat a_1^\dagger \hat\sigma_{2e} +i
 \hat\sigma_{e1} g_2\hat a_2,
\label{eq:aprop}
\end{alignat}
where we have introduced the inversion operators
$\hat\sigma_{zi}=\hat\sigma_{ee}-\hat\sigma_{ii}$. Without loss of generality, we assume that
$g_i$ is real, as we can always displace the origin of the
electromagnetic field phase to absorb any phase factor of the coupling constant into the
definition of the operators $\hat a_i$. The $\hat f_i$ terms are
delta-correlated collective Langevin operators, which account for the
noise introduced by the coupling between the atomic system to the free
radiation field. These operators have vanishing mean values and
correlation functions of the form
\begin{equation}
\langle \hat f_{x}(z,t)
\hat f_{y}(z,t')\rangle=\frac{L}{N}D_{xy}\delta(t-t')\delta(z-z')\, ,
\end{equation}
where the diffusion coefficients $D_{xy}$ can be obtained from the
generalized Einstein equations.
Note that Eq.~(\ref{eq:eprop}) is the extension to the quantum domain
of the classical wave equation in the slowly varying field approximation~\cite{lb:scully}.
This equation describes the propagation of the fields through the atomic ensemble with
the atomic dipoles acting as the driving terms.
A detailed derivation of
Eqs.~(\ref{eq:eprop})-(\ref{eq:aprop}) is given in~\cite{rv:richter}.

\section{Tri-partite entanglement with continuous variables}
\label{sc:entanglement}
The Hilbert space of the system is $H=H_1\otimes H_2 \otimes H_A$
where $H_j$ is the Hilbert space of  field $j$ and $H_A$ is
the Hilbert space of the atoms. The state of the system, given by the density
operator $\rho:H\rightarrow H$, is entangled if it cannot be written
as
\begin{equation}
  \label{eq:separable}
  \rho=\sum_i p_i \rho_1^{(i)}\otimes\rho_2^{(i)}\otimes\rho_A^{(i)}\, ,
\end{equation}
where $\rho_1^{(i)}:H_1\rightarrow H_1$, $\rho_2^{(i)}:H_2\rightarrow H_2$,
$\rho_A^{(i)}:H_A\rightarrow H_A$. It is however possible for a multi-partite entangled state to be partially separable~\cite{RevModPhys.81.865}. For example, consider the state
\begin{eqnarray}
  \label{eq:tripartite}
  \rho&=&p_1\rho_{12}\otimes\rho_A\,+\,p_2\rho_{1A}\otimes\rho_2+p_3\rho_{2A}\otimes\rho_1\, ,
\end{eqnarray}
where $\rho_{12}:H_1\otimes H_2\rightarrow H_1\otimes H_2$,
$\rho_{1A}:H_1\otimes H_A\rightarrow H_1\otimes H_A$,
$\rho_{2A}:H_2\otimes H_A\rightarrow H_2\otimes H_A$.
Equation~(\ref{eq:tripartite}) cannot necessarily  be written as
Eq.~(\ref{eq:separable}) and therefore can be entangled. However, given that Eq.~(\ref{eq:tripartite}) represents a statical mixture of states with bi-partite entanglement, it does not represent a genuine tri-partite entangled state.

In order to study the presence of genuine tri-partite entanglement between
the two fields, probe and control, and the atoms, we use
an approach based on separability conditions for continuous variables
\cite{van_loock_detecting_2003,PhysRevA.90.062337}. To apply such conditions to our
system, three pairs of operators are needed, one for each of the three
subsystems, with each pair having a position-momentum commutation
relation. For the fields we consider the generalized quadrature operator
\begin{equation}
  \hat Y^\theta_j(z,\omega)=\hat a_j(z,\omega) \exp(-i\theta) +  \hat a_j^\dagger(z,\omega) \exp(i\theta)\, ,
\end{equation}
with $j=1,2$ and where
$\hat a_j(z,\omega)$ is the annihilation operator of a photon of
frequency $\omega$ for field $j$. With this definition for the generalized quadrature operator the commutation relation takes the form
$[\hat Y^{\theta}_j(z,\omega),\hat Y^{\theta+\pi/2}_j(z,\omega)]=2 i$, which leads to a  quadrature noise (variance) for a coherent state of 1. For
the atomic medium the choice is not as clear as the atomic operators do not satisfy the required commutation relations.  Instead, we introduce the following approach, we denote the electron position operator as $\hat x^{(s)}$ and  its momentum operator as $\hat p^{(s)}$. Using equations analogous to  Eq.~(\ref{eq:oz}), we
construct continuous versions of the position, $\hat x(z)$,  and momentum, $\hat p(z)$, operators that satisfy $[\hat x(z),\hat p(z)]=i\hbar$.  We then define unitless operators analogous to the field quadratures as
\begin{eqnarray}
\hat X(z)&=&\sqrt{2 \omega_a m}\,\hat x(z)/\sqrt{\hbar}\nonumber\, ,\\
\hat P(z)&=&\sqrt{2}\,\hat p(z)/\sqrt{\omega_a\, m\hbar}\, ,\label{eq:X_P_def}
\end{eqnarray}
where $m$ is the mass of the electron. These operators satisfy the necessary commutation relation, $[\hat X(z),\hat P(z)]=2 i$.

With this approach we are left with the following operators
\begin{equation*}
\hat Y_1^0(z,\omega),\hat Y_1^{\pi/2}(z,\omega),\hat
Y_2^0(z,\omega),\hat Y_2^{\pi/2}(z,\omega),\hat X(z),\hat P(z)\, ,
\end{equation*}
that form three pairs, one for each subsystem, of continuous operators with position-momentum
commutation relations. With these operators we can derive the following set of inequalities analogous to the ones in~\cite{PhysRevLett.97.140504}
\begin{subequations}\label{eq:vanloock_inequalities}
  \begin{eqnarray}
  I_1(z)&=&\langle \Delta^2(\hat Y_1^{\pi/2}(z,\omega)-\hat Y_2^{\pi/2}(z,\omega))\rangle+\langle
         \Delta^2(\hat Y_1^{0}(z,\omega)+\hat
            Y_2^{0}(z,\omega))\rangle\geq 4\, ,\label{eq:vanloock_inequalities_a}\\
  I_2(z)&=&\langle \Delta^2(\hat P(z)+\hat Y_1^{\pi/2}(z,\omega))\rangle+\langle
         \Delta^2(\hat Y_1^{0}(z,\omega)+h_2 \hat
            Y_2^{0}(z,\omega)-\hat X(z))\rangle\geq 4\, ,\label{eq:vanloock_inequalities_b}\\
  I_3(z)&=&\langle \Delta^2(\hat P(z)+\hat Y_2^{\pi/2}(z,\omega))\rangle+\langle
         \Delta^2(h_1 \hat Y_1^{0}(z,\omega)+ \hat
            Y_2^{0}(z,\omega)-\hat X(z))\rangle\geq 4\, ,\label{eq:vanloock_inequalities_c}
\end{eqnarray}
\end{subequations}
where $h_i$ are arbitrary real parameters. Violation of each of these
inequalities provides information about the quantum nature of the
combined probe, control, and atomic systems. If
Eq.~(\ref{eq:vanloock_inequalities_a}) is not satisfied, then the two
fields, control and probe, are entangled. This inequality corresponds to the Duan
criteria for inseparability of two optical fields~\cite{PhysRevLett.84.2722}. If
Eq.~(\ref{eq:vanloock_inequalities_b}) is not satisfied, then the control field, $\hat{E}_1$,
is entangled with the atoms. Finally, if
Eq.~(\ref{eq:vanloock_inequalities_c}) is not satisfy, then the probe field, $\hat{E}_2$
is entangled with the atoms. For a pure state a simultaneous violation of these three inequalities indicates the presence of genuine tri-partite entanglement in the system.  However, for the case of a mixed state, as the one given in Eq.~(\ref{eq:tripartite}), this is not necessarily the case. In this case, to guarantee the presence of tri-partite
genuine entanglement, a more stringent condition given by the sum of the three
inequalities
\begin{equation}
  \label{eq:genuine_entanglement_inequalities}
  \sum_{i=1}^3 I_i<4\, ,
\end{equation}
has to be satisfied~\cite{PhysRevA.90.062337}.

\section{Results}
\label{sc:results}
To solve the set of Eqs.~(\ref{eq:eprop}) and (\ref{eq:aprop}), we
transform them into a set of equivalent $c$-number stochastic differential
equations that can be used to calculate the spectra of the product of two normally ordered operators (see
\cite{rv:davidovich} for a detailed description of this
transformation). We then solve these $c$-number differential equations in the small noise approximation and in the
stationary regime (see \cite{rv:pablomarc} for details for the $\Lambda$ system and \cite{rv:pablomarc2,rv:richter} for the values of
the diffusion coefficients). With this approach we can obtain
the mean value of any operator that appears in the equations as well as the
spectrum of normally ordered operator products.
The normal order we use is
\begin{equation}\label{eq:order}
\hat a_{2}^{\dagger }, \hat a_{1}^{\dagger },
{\hat\sigma}_{e2},{\hat\sigma}_{e1},{\hat\sigma}_{12},\hat\sigma_{z1},\hat\sigma_{z2},\hat\sigma_{21},\hat\sigma_{1e},
\hat\sigma_{2e}, \hat a_1, \hat a_2\, .
\end{equation}
To relate solutions of the $c$-number differential equations to the spectrum, recall the definition of the spectrum in the stationary regime
\begin{equation}
  \langle  \delta \hat  o_1(\omega) \delta \hat o_2(\omega)\rangle=
  \lim_{\tau\rightarrow\infty}\int\limits_{-\infty}^\infty
  e^{-i\omega \tau} \langle \delta \hat o_1(t+\tau)\delta \hat
  o_2(t)\rangle d\tau\,
\label{eq:spec_gen}
\end{equation}
where
\begin{equation}
\delta \hat o(t)=\hat o(t) -\langle \hat o(t)\rangle\, ,
\end{equation}
with $\hat o_i$ being any system operator. Our solution to the
system of equations allows us to calculate the spectrum when $\hat o_1$
appears on the left of $\hat o_2$ in the list given in
(\ref{eq:order}). For example, we can directly calculate
$\langle{\hat\sigma}_{12}(\omega)\hat a_1(\omega)\rangle$. However, if we want
to calculate
$\langle\hat\sigma_{2e}(\omega){\hat\sigma}_{e1}(\omega)\rangle$ we need to use the commutation  relation between $\hat\sigma_{2e}(\omega)$
and ${\hat\sigma}_{e1}(\omega)$ to first rewrite the expression in normal order.

We are now in a position to calculate any observable that can be written
as a sum of normally ordered products of two elements in the list
given in (\ref{eq:order}). We focus on the noise spectra of the
field quadratures and atomic observables in order to evaluate the inequalities given in Eqs.~(\ref{eq:vanloock_inequalities}) and look for conditions for which all three inequalities are simultaneously violated and Eq.~(\ref{eq:genuine_entanglement_inequalities}) is satisfied.

As an input into the atomic medium we consider entangled twin beams (two-mode squeezed states) with one of the fields acting as the control field and the other one as the probe field. More specifically, we consider the field boundary conditions at the input of the atomic medium, $z=0$, to be given by
\begin{eqnarray*}
  \langle \delta \hat a_i^\dagger(\omega) \delta \hat
  a_i(\omega)\rangle&=&\eta  \sinh ^2(r)\, ,\\
  \langle \delta \hat a_1 (\omega)\delta \hat
  a_2(\omega)\rangle&=&\langle \delta \hat a_2 (\omega)\delta \hat
                        a_1(\omega)\rangle=-\eta  \cosh (r) \sinh
                        (r)\, ,
\end{eqnarray*}
with $\langle \hat a_1(\omega)\rangle=\langle \hat a_2(\omega)\rangle=0$ when $\omega\neq \omega_a$,
$\langle \hat a_1(\omega_a)\rangle=\alpha_1$, and
$\langle \hat a_2(\omega_a)\rangle=\alpha_2$.  Without loss of generality we can assume
$\alpha_i\in\mathbb{R}$. In these expressions $r$ is the squeezing parameter, with $r=0$ corresponding to a coherent state, and $\eta$ represents the
losses in the squeezed state preparation, which are taken to be the same for both fields.  This input optical state corresponds to bright or displaced two-mode squeezed states (bright twin beams) with a carrier frequency on resonance with the atomic transition, $\omega_{a}$. Thus,  the sidebands ($\omega\neq \omega_a$) of the initial state are in a vacuum two-mode squeezed state, while the carrier ($\omega=\omega_a$) is in a displaced two-mode squeezed
state. The input fields drive the two dipolar transitions of
the atoms in resonance, creating the conditions for EIT. When the system reaches its
stationary state
$\langle \hat \sigma_{ei}\rangle=\langle \hat\sigma_{ie}\rangle=0$
and
$\langle\hat
\sigma_{21}\rangle=-g_1g_2\Omega_1\Omega_2/(g_1^2\Omega_1^2+g_2^2\Omega_2^2)$,
where we have introduced the Rabi frequency $\Omega_i=g_i\alpha_i$. Such expectation values are a signature of coherence between the atomic ground states $|1\rangle$ and
$|2\rangle$.

\subsection{Control and probe field quadratures}
We first consider how the noise properties of the individual fields change as they propagate through the atomic medium. In the ideal case of no decoherence, as we show, changes to the correlations of and between the fields can happen even when the field mean values remain constant given that the system is under EIT conditions.
To study the evolution of the quantum properties of the fields as they propagate through the medium, we consider the spectrum of the quadratures, $i=j$, and their cross-correlations,
$i\neq j$,
\begin{equation}
{\mathcal S}_{ij}^{\theta_1,\, \theta_2}(\omega,z) = \langle \delta Y_i^{\theta_1}(\omega,z)\delta Y_j^{\theta_2}(\omega,z)\rangle\, ,
\label{eq:spec}
\end{equation}
which we calculate after solving for the $c$-number stochastic differential equations using the method described above. The most
general analytical result for Eq.~(\ref{eq:spec}) is given in the Appendix.

We now specialize to some particular cases. First we consider the limit in which the control field is much stronger than the probe field, $\Omega_1\gg\Omega_2$, and $g_1=g_2$, such that in steady state all the atomic population is in level $\ket{2}$. Under these conditions
\begin{equation}\label{eq:alpha0pump}
{\mathcal S}_{11}^{\theta_1=0,\,
  \theta_2=0}(\omega,z)={\mathcal S}_{11}^{\theta_1=0,\,
  \theta_2=0}(\omega,z=0)
\end{equation}
and
  \begin{equation}\label{eq:alpha0}
{\mathcal S}_{22}^{\theta_1=0,\,
  \theta_2=0}(\omega,z)=1+\eta\left(\frac{e^{-2 r}}{2}+\frac{e^{2
      r}}{2}-1\right) e^{-2 Q_a(\omega) z}\, ,
\end{equation}
where $Q_a(\omega)$ is given in the Appendix and shown in Fig.~\ref{fig:absorption}.
Equation~(\ref{eq:alpha0pump}) shows that the noise properties of the quadrature of the control field are not affected by the atoms. On the other hand, Eq.~(\ref{eq:alpha0}) shows that the quadrature fluctuations of the probe field, which by itself has thermal statistics, tend towards a coherent state as the fields propagate.  The length scale, $z_l\approx 1/Q_a(\omega)$, for this
to happen depends on the field frequency $\omega$ and the coupling strength
between the field and the atoms. As can be concluded from Fig.~\ref{fig:absorption}, $z_l$ tends to infinity at the carrier frequency ($\omega=\omega_{a}$) and is minimum for sidebands at $\omega=\omega_a\pm \Omega_1/2$, for which $Q_a(\omega)$ is maximum. The corresponding equations for the quadrature $\theta=\pi/2$ can be obtained by changing the sign of $r$ in Eqs.~(\ref{eq:alpha0pump}) and (\ref{eq:alpha0}). Thus, for this case the behavior of both quadratures is the same.
\begin{figure}
\centering
  \includegraphics[width=7cm]{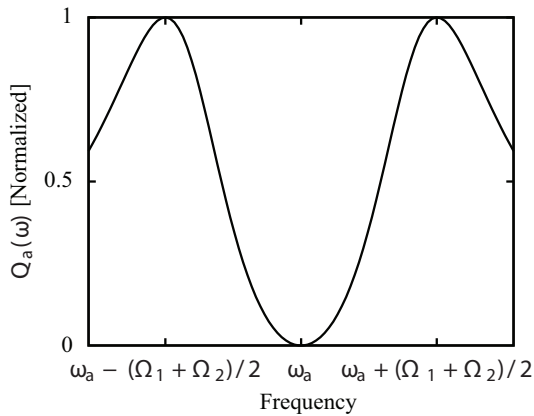}
  \caption{Parameter $Q_a(\omega)$ as a function of frequency. The length scale over which the quadrature noise of the fields is modified
    depends on its frequency $\omega$ and is given by the inverse of parameter
    $Q_a(\omega)$.}
  \label{fig:absorption}
\end{figure}

\begin{figure}
\centering
  \includegraphics[width=7cm]{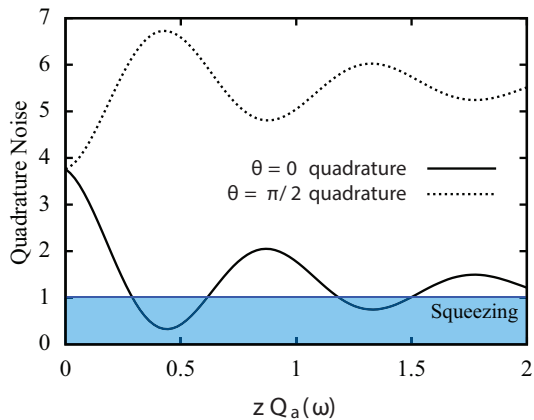}
  \caption{Single field quadrature noise for the probe and control fields as a function of position in the atomic medium for $\Omega_2=(1+\sqrt{2})\Omega_1$. Given the definition of the quadratures, a value of 1 corresponds to the noise level of a coherent state. Note that for both quadratures the noise oscillates as the fields propagate through the medium. Regions below the blue line (shaded blue region) indicate the presence of squeezing for the $\theta=0$ quadrature. We consider the case for which $\eta=1$, $r=1$, and $Q_o/Q_a=7$.}
  \label{fig:quadratures_oscillation}
\end{figure}

Next, we consider the case for which $\Omega_2=\Omega_1(1\pm \sqrt{2})$ and the limit in which the exponential decay due to $Q_a$ can be neglected.  In this case
\begin{equation}\label{eq:oscillations}
  {\mathcal S}_{ii}^{\theta_1=0,\,
  \theta_2=0}(\omega,z)=1+\frac{\eta}{4} \left(-e^{-2 r}+1\right)
                          \left[\left(e^{2
                                         r}+1\right) \cos (Q_o z)+e^{2
                                         r}-3\right]\, ,
\end{equation}
and
\begin{equation}\label{eq:oscillations11}
  {\mathcal S}_{ii}^{\theta_1=\pi/2,\,
  \theta_2=\pi/2}(\omega,z)=1+\frac{\eta}{4} \left(-e^{2 r}+1\right)
                           \left[\left(e^{-2
                                         r}+1\right) \cos (Q_o z)+e^{-2
                                         r}-3\right]\, ,
\end{equation}
with $Q_o(\omega)$ given in the Appendix and $i=1,2$. For these parameters, the
noise of the quadratures of control and probe fields are equal to each other.
Equation~(\ref{eq:oscillations}) shows that the single beam noise for quadrature $\theta=0$ oscillates as the field propagates through the medium between excess noise,
$1+\frac{1}{2} \left(e^{r}-e^{-r}\right)^2$, and single-mode squeezing,
$e^{-2 r}$ (assuming no losses in the two-squeezed state preparation,
$\eta=1$). The single beam noise for quadrature
$\theta=\pi/2$ also shows oscillations, as can be seen from  Eq.~(\ref{eq:oscillations11}), but always exhibits excess noise.  The frequency of these oscillations is given by $Q_o(\omega)$
and is different for each sideband of the field due to the frequency dependence of $Q_o(\omega)$. Note that the relation between the chosen values of $\Omega_i$ maximizes $Q_o(\omega)$ and thus the frequency of the oscillations. As these expressions show, the noise properties of
the field can change between two extrema without the mean value of the
field being altered. For the case in which $Q_a(\omega)$ can not be neglected, the oscillations decays as
a function of $z$, as can be seen in
Fig.~\ref{fig:quadratures_oscillation}.

While an oscillatory behavior of the noise properties of the single field quadratures of
the probe and control  under EIT conditions had been previously reported for the
case of initially separable fields~\cite{rv:pablomarc2}, such behavior was only observed when the initial quadrature noise spectra of the fields were different. For separable input fields, the oscillations disappear when the noise spectra of the quadratures for the control and probe fields become the same. Our results show that for the case of initially entangled probe and control fields this situation changes, as oscillations are observed even for the case of equal initial noise spectra of the quadratures of the fields.

\begin{figure}
\centering
  \includegraphics[width=7cm]{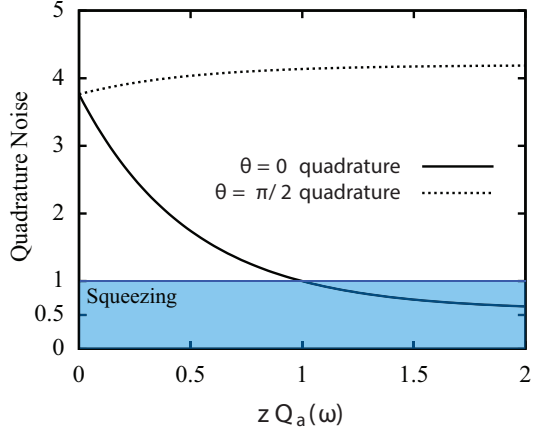}
  \caption{Single field quadrature noise for the probe and control fields as a function of position in the atomic medium for $\Omega_1=\Omega_2$.  Given the definition of the quadratures, a value of 1 corresponds to the noise level of a coherent state. Note that the
    $\theta=0$ quadrature for each field becomes squeezed (shaded blue region) as the fields propagate through the medium. Parameters: $\eta=1$ and $r=1$.}
  \label{fig:quadratures}
\end{figure}

Finally, we consider the case when $\Omega_1=\Omega_2=\Omega$, for which we obtain
\begin{equation}\label{eq:qudratures_equal_1}
  {\mathcal S}_{ii}^{\theta_1=0,\,
    \theta_2=0}(\omega,z)=1+\eta\left[\frac{1}{2}   \left(e^{2
        r}-1\right) e^{-2 Q_a z}+\frac{1}{2}   \left(e^{-2
        r}-1\right)\right]\, ,
\end{equation}
and
\begin{equation}\label{eq:qudratures_equal_2}
  {\mathcal S}_{ii}^{\theta_1=\pi/2,\,
    \theta_2=\pi/2}(\omega,z)=1+\eta\left[\frac{1}{2}   \left(e^{2
        r}-1\right) +\frac{1}{2}  \left(e^{-2
        r}-1\right) e^{-2 Q_a z}\right]\, ,
\end{equation}
with $i=1,2$.
For an initial pure state ($\eta=1$) with $r>0$, before propagation
through the medium, $z=0$, the individual quadratures for both control
($i=1$) and probe ($i=2$) fields show excess noise with respect to a
coherent state. As the fields propagate through the medium, the noise
spectra for the $\theta=0$ and $\theta=\pi/2$ quadratures as a
function of  position $z$ evolve as shown in
Fig.~\ref{fig:quadratures}. As can be seen, the effect of EIT on the
fields is to squeeze the $\theta=0$ quadrature. The length scale,
$z_l$, for this to happen is given by $z_l\approx 1/Q_a(\omega)$ and
is minimum when $Q_a(\omega)$ is maximum, which happens when
$\omega=\omega_a\pm \Omega$ as can be seen in
Fig.~\ref{fig:absorption}. Note that in general the actual quadrature that is
squeezed depends on the argument of $g_i$, which, as described in
section~\ref{sc:equations}, is absorbed in the definition of the field
operators.

In the limit $z\rightarrow \infty$ for sidebands around the carrier ($\omega\neq \omega_a$), Eqs.~(\ref{eq:qudratures_equal_1}) and (\ref{eq:qudratures_equal_2}) reduce to
\begin{eqnarray}
{\mathcal S}_{ii}^{\theta_1=0,\, \theta_2=0}(\omega)&=&\frac{1}{2}+\frac{e^{-2r}}{2}\, ,\\
  {\mathcal S}_{ii}^{\theta_1=\pi/2,\, \theta_2=\pi/2}(\omega)&=&\frac{1}{2}+\frac{e^{2r}}{2}\, ,
\end{eqnarray}
which represent squeezing for the $\theta=0$ quadrature and excess noise for the
$\theta=\pi/2$ quadrature. Note that while squeezing is present, the field is not a minimum uncertainty state.

\subsection{Atomic dipole fluctuations}

We next consider the evolution of the atomic medium as the fields
propagate through it. To obtain the spectra of the fluctuations for
$\hat X$ and $\hat P$ (see Eqs.~(\ref{eq:X_P_def})), we write these
operators as a function of  atomic operators. The electron position
operator of atom $s$, $\hat x^{(s)}$, can be written as
\begin{eqnarray}
  \label{eq:position_operator}
  \hat x^{(s)}&=&\hat J_{+x}^{(s)}+\hat J_{-x}^{(s)}\, ,\\
  \hat J_{+x}^{(s)}&=&\sum_{i>j} x^{(s)}_{ij}|i\rangle^{(s)}{}^{(s)}\langle j|=\hat J_{-x}^{(s)\dagger}\, ,
\end{eqnarray}
where $|i\rangle^{(s)}$ is an eigenstate of the atomic Hamiltonian,
belonging to the Hilbert space of atom $s$, with label $i$, and
$x^{(s)}_{ij}={}^{(s)}\langle i|\hat x|j\rangle^{(s)}$. Its momentum
operator, $\hat p^{(s)}$, can be written as a function of
$x_{ij}^{(s)}$ as
\begin{eqnarray}\label{eq:momentum_operator}
  \hat p^{(s)}&=&\hat J_{+p}^{(s)}-\hat J_{-p}^{(s)}\\
  \hat J_{+p}^{(s)}&=&\sum_{i>j} p^{(s)}_{ij}|i\rangle^{(s)}{}^{(s)}\langle j|\nonumber\\
          &=&\sum_{i>j} i\,m\,(\omega_i-\omega_j)x^{(s)}_{ij}|i\rangle^{(s)}{}^{(s)}\langle j|=\hat J_{-p}^{(s)\dagger},
\end{eqnarray}
where $p^{(s)}_{ij}={}^{(s)}\langle i| \hat p|j\rangle^{(s)}$.

If we consider a three level atom with $x^{(s)}_{1e}=x^{(s)}_{2e}=x$, $\omega_e-\omega_{1}=\omega_e-\omega_{2}=\omega_a$ we can then write the continuous position and momentum operators as
\begin{eqnarray}
  \label{eq:x_p_approx}
  \hat{x}(z)&=&\hat J_{+x}(z)+\hat J_{-x}(z)\approx x[\hat J_+(z)+\hat J_-(z)]\, ,\\
  \hat{p}(z)&=&\hat J_{+p}(z)-\hat J_{-p}(z)\approx i\, m \,\omega_a \, x \, [\hat J_+(z)-\hat J_-(z)]\, ,
\end{eqnarray}
with $\hat J_+(z)=\hat \sigma_{e1}(z)+\hat \sigma_{e2}(z)=\hat J_-^\dagger(z)$. Using Eqs.~(\ref{eq:x_p_approx}) we can then write Eqs.~(\ref{eq:X_P_def})
as
\begin{subequations}\label{eq:X_P_simp}
  \begin{eqnarray}
    \hat X&=&\sqrt{2\omega_a m}x[\hat J_+(z)+\hat J_-(z)]/\sqrt{\hbar}\,,\nonumber\\
    \hat P&=&i\sqrt{2\omega_a m}x[\hat J_+(z)-\hat J_-(z)]/\sqrt{\hbar}\,.
  \end{eqnarray}
\end{subequations}
This allows us to obtain
\begin{equation}
\langle\mathop{:}\nolimits\Delta^2\hat X(\omega)\mathop{:}\nolimits\rangle=
\frac{2\omega_am x^2}{\hbar}\sum^2_{i=1,j=1}\Big
                                       [\hat\sigma_{ei}(\omega)\hat\sigma_{ej}(\omega)+\hat\sigma_{ie}(\omega)\hat\sigma_{je}(\omega)+2\hat\sigma_{ei}(\omega)\hat\sigma_{je}(\omega)\Big],\nonumber\\\label{eq:delta_x}
\end{equation}
\begin{equation}
\langle\mathop{:}\nolimits\Delta^2\hat P(\omega)\mathop{:}\nolimits\rangle=
-\frac{2\omega_am x^2}{\hbar}\sum^2_{i=1,j=1}\Big
                                       [\hat\sigma_{ei}(\omega)\hat\sigma_{ej}(\omega)+\hat\sigma_{ie}(\omega)\hat\sigma_{je}(\omega)-2\hat\sigma_{ei}(\omega)\hat\sigma_{je}(\omega)\Big],\nonumber\\\label{eq:delta_p}\,
\end{equation}
where $\mathop{:}\nolimits\hat o\mathop{:}\nolimits $ indicates normal ordering.

For the case in which $\Omega_1=\Omega_2$, we obtain that (see section~\ref{sc:results})
\begin{equation}\label{eq:XP_fluctuation_zero}
\langle\mathop{:}\nolimits\Delta^2\hat P(\omega)\mathop{:}\nolimits\rangle=\langle\mathop{:}\nolimits\Delta^2\hat X(\omega)\mathop{:}\nolimits\rangle=0\, ,
\end{equation}
and
\begin{equation}
\langle \mathop{:}\nolimits\Delta^2 \hat J_+\mathop{:}\nolimits \rangle=\langle \mathop{:}\nolimits\Delta^2 \hat J_-\mathop{:}\nolimits
\rangle=0\, .
\end{equation}
To understand the implications of this result, let us consider the random process
$\sigma_{e1}(t)+\sigma_{e2}(t)-\langle
\sigma_{e1}+\sigma_{e2}\rangle$, which corresponds to the fluctuation
around its mean value of $J_+$ and where we have used the nomenclature $o$ for the $c$-number random variable associated with operator
$\hat o$ produced by the transformation from stochastic operator
equations to the equivalent set of $c$-number stochastic equations. Now, if
$\langle \mathop{:}\nolimits\Delta^2 \hat J_+\mathop{:}\nolimits
\rangle=0$ then
$\sigma_{e1}(t)+\sigma_{e2}(t)-\langle
\sigma_{e1}+\sigma_{e2}\rangle=0$ for each realization of the
stationary stochastic process. This implies that the noise from each of the
dipoles in the $\Lambda$ system cancels each other out. It is thus reasonable to expect strong correlations between each of the fields and the atoms, as the fluctuations of both dipoles, each of them driven by one of the
fields, are correlated in such a strong way that when excess noise
(with respect to the mean) appears in one dipole transition, a
corresponding reduction in the noise appears in the
other. As the dipoles themselves act as sources for the fields, the fields are modified accordingly with the effect that the noise spectra of the quadratures of the control and probe fields are the same as they propagate through the atomic medium (see
Eqs.~(\ref{eq:qudratures_equal_1}) and (\ref{eq:qudratures_equal_2})).
In the next section, we show that these correlations are strong enough
to lead to the presence of genuine tri-partite entanglement between
the two fields and the atoms.

We also calculate the noise spectrum for the atomic coherence between
the two ground states. It has been previously shown that when the initial
state is a separable state with the probe field in a squeezed vacuum state
and the control field in a coherent state, the ground state atomic coherence
becomes squeezed as a result of the EIT process~\cite{rv:dantan4}. Nevertheless, when the initial
condition is a two-mode squeezed state, spin squeezing does not result from the
interaction. Note that as opposed to Ref.~\cite{rv:dantan4}, for the two-mode squeezed state
the quadrature noise of the fields entering the medium is the same for
the probe and control fields and is larger than the one of the coherent state.

\subsection{Entanglement}

We can now consider the entanglement properties of the system by using the above results to calculate the inequalities given by Eqs.~(\ref{eq:vanloock_inequalities}). Through the use of Eqs.~(\ref{eq:X_P_simp}) and the commutation relations between
$\hat J_+$ and $\hat J_-$, we can write Eqs.~(\ref{eq:vanloock_inequalities}) as
\begin{subequations}    \label{eq:vanloock_inequalities_no}
    \begin{eqnarray}
        I_1(\omega)&=&\langle :\Delta^2(\hat Y_1^{\pi/2}(\omega)-\hat Y_2^{\pi/2}(\omega)):\rangle+\langle
                     :\Delta^2(\hat Y_1^{0}(\omega)+\hat Y_2^{0}(\omega)):\rangle\geq 0\, ,\label{eq:vanloock_inequalities_no_a}\\
      I_2(\omega)&=&\langle: \Delta^2(\hat P+\hat Y_1^{\pi/2}(\omega)):\rangle+\langle:
                     \Delta^2(\hat Y_1^{0}(\omega)+h_2 \hat Y_2^{0}(\omega)-\hat X):\rangle\geq 1-\frac{4 m \omega_a
                     x^2}{\hbar}\, ,\label{eq:vanloock_inequalities_no_b}\\
      I_3(\omega)&=&\langle: \Delta^2(\hat P+\hat Y_2^{\pi/2}(\omega)):\rangle+\langle:\Delta^2(h_1
                     \hat Y_1^{0}(\omega)+ \hat Y_2^{0}(\omega)-\hat X):\rangle\geq 1-\frac{4 m \omega_a
                     x^2}{\hbar}\, ,\label{eq:vanloock_inequalities_no_c}
    \end{eqnarray}
\end{subequations}
and Eq.~(\ref{eq:genuine_entanglement_inequalities}) as
\begin{equation}
  \label{eq:genuine_entanglement_no}
    \sum_{i=1}^3 I_i<-6-2\frac{4 m \omega_a
                     x^2}{\hbar}\, .
\end{equation}
From the solution of the system equations when $\Omega_1=\Omega_2$, we get that
\begin{subequations}
  \begin{equation}
    \label{eq:s1_full}
    I_1=\left(2 e^{-2  r}-2\right) e^{-2 Q_a z}+2 e^{-2 r}-2\, ,
  \end{equation}
  \begin{equation}
    \label{eq:s2_full}
    I_2=\left(\frac{1}{2} h_2^2 e^{2 r}-\frac{h_2^2}{2}+h_2 e^{2 r}-h_2+\cosh{2 r}-1\right) e^{-2 Q_a
           z}+\frac{1}{2} h_2^2 e^{-2 r}-\frac{h_2^2}{2}-h_2 e^{-2 r}+h_2+\cosh{2 r}-1\, ,
  \end{equation}
  \begin{equation}
    \label{eq:s3_full}
    I_3=\left(\frac{1}{2} h_1^2 e^{2 r}-\frac{h_1^2}{2}+h_1 e^{2 r}-h_1+\cosh{2 r}-1\right) e^{-2 Q_a
           z}+\frac{1}{2} h_1^2 e^{-2 r}-\frac{h_1^2}{2}-h_1 e^{-2 r}+h_1+\cosh{2 r}-1\, .
  \end{equation}
\end{subequations}
It is possible to see from Eq.~(\ref{eq:s1_full}) that when $r>0$ then
$I_1(z)<0$ for all $z$, which implies that the two fields remain
entangled as they propagate through the atomic medium. On the other
hand, violations of Eqs.~(\ref{eq:vanloock_inequalities_no_b}) and
(\ref{eq:vanloock_inequalities_no_c}) depend on the values of $r$,
$h_1$, $h_2$ and $z$.

\begin{figure}
\centering
  \includegraphics[width=7cm]{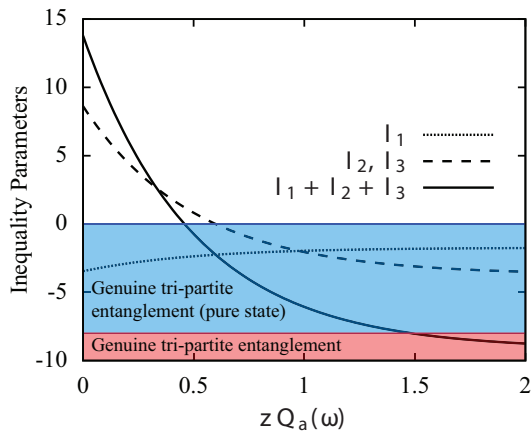}
  \caption{Inequality parameters as a function of position
    inside the atomic medium. For the case of a pure state, when $I_1$ (dotted line) and $I_2=I_3$ (dashed line) are below zero (blue shaded region) the system exhibits genuine tri-partite entanglement.
    In general, the more stringent condition for genuine tri-partite entanglement given by considering $I_1+I_2+I_3$ (solid line) is also satisfied (red shaded region). Parameters:
    $r=1$, $h_1=h_2=-3$, and $\eta=1$.}
  \label{fig:Inequalities}
\end{figure}

Figure~\ref{fig:Inequalities} shows an example
of the evolution of the inequality parameters, $I_i$ for $i=1,2,3$, as a function of position $z$ in
the atomic medium. As the fields propagate, $I_1$ increases from its
initial value; however $I_2$ and $I_3$ decrease. This behavior shows that the initial entanglement between the fields is redistributed to entanglement between each of the fields and the atomic ensemble. As can be seen from the blue shaded region, for the case of pure states ($\eta=1$), the system evolves into a genuine tri-partite entangled state for $z\,Q_{a}$ slightly larger than $0.5$. Furthermore, as the fields continue to propagate through the medium, the entanglement becomes stronger such that when $z\,Q_{a}>1.5$ the more stringent and general condition for genuine tri-partite entanglement, $\sum_i I_i<-8$ (taking into account that $4 m \omega_a x^2/\hbar\sim 1$), is also satisfied.
In general, it is possible to find values of $h_i$ for which
Eq.~(\ref{eq:genuine_entanglement_no}) is violated when $z$ satisfy the condition
\begin{equation*}
2 - \cosh(2 r) + \cosh[2 Q_a(\omega) z]>0\, .
\end{equation*}
In the limit $r\gg 1$, this condition is satisfied when $z>r/Q_a(\omega)$, which shows that the distance at which genuine tri-partite entanglement is present for a sideband at frequency $\omega$ is minimum when $Q_a(\omega)$ is maximum and that entanglement is not generated at the on-resonance carrier frequencies.

These results show that even though there is no initial entanglement between the atoms and the fields, as the fields propagate through the
medium their interaction creates a state with genuine tri-partite
entanglement. Thus, the medium composed of $\Lambda$ atoms in an EIT configuration effectively
redistributes the initial two-mode entanglement between the fields to
genuine tri-partite entanglement between the two fields and the atomic ensemble.

\section{Measuring genuine tri-partite atom-field entanglement}\label{sc:measuring}

We now propose a technique to measure the noise properties of the field and atom operators needed to evaluate Eqs.~(\ref{eq:vanloock_inequalities_no}) using only field observables. We specialize to the case described above in which the two fields drive the atoms with the
same strength, $\Omega_1=\Omega_2$. Given that Eq.~(\ref{eq:vanloock_inequalities_no_a}) consists only of field observables, its evaluation is experimentally accessible through measurements of the quadrature noise
spectra for each of the fields and the correlations between them~\cite{RevModPhys.81.1727,Boyer_2008,Dowran_2018}.
Equations~(\ref{eq:vanloock_inequalities_no_b}) and (\ref{eq:vanloock_inequalities_no_c}), on the other hand, contain both atomic and field observables, which makes it hard to evaluate them directly. It is however possible to find measurements on the fields  to estimate both $I_{2}$ and $I_{3}$. To show that this is the case, we take advantage of the fact that Eq.~(\ref{eq:XP_fluctuation_zero}) implies that
\begin{equation}\label{eq:noise_is_zero}
\delta\sigma_{1e}+\delta\sigma_{2e}\pm(\delta\sigma_{e1}+\delta\sigma_{e2})=0.\,
\end{equation}
This makes it possible to rewrite $I_2(\omega)$ and $I_3(\omega)$ as
\begin{subequations}
  \begin{eqnarray}
    I_2(\omega)&=&\langle: \Delta^2(\hat Y_1^{\pi/2}(\omega)):\rangle+\langle:
                   \Delta^2(\hat Y_1^{0}(\omega)+h_2 \hat Y_2^{0}(\omega)):\rangle\, ,\label{eq:vanloock_inequalities_no_b_2}\\
    I_3(\omega)&=&\langle: \Delta^2(\hat Y_2^{\pi/2}(\omega)):\rangle+\langle:\Delta^2(h_1
                   \hat Y_1^{0}(\omega)+ \hat Y_2^{0}(\omega)):\rangle\, ,\label{eq:vanloock_inequalities_no_c_2}
  \end{eqnarray}
\end{subequations}
which become experimentally accessible as both expressions now depend only on the
quadrature noise spectra for each field and their correlations. In order
to find field measurements that imply that
Eq.~(\ref{eq:noise_is_zero}) is satisfied, we take advantage of the fact that substituting
$\hat a_j=\delta \hat a_j+\langle \hat a_j \rangle$ into
Eq.~(\ref{eq:eprop}), summing the equations for $\hat a_1$,
$\hat a_2$, $\hat a^\dagger_1$, $\hat a^\dagger_2$, and transforming
the operator equations to $c$-number equations results in
\begin{equation}
\left(\frac{\partial}{\partial t}+c\frac{\partial}{\partial
    z}\right) \delta ( Y_1^0+ Y_2^0) = -i g N[\delta\sigma_{1e}+\delta\sigma_{2e}+(\delta\sigma_{e1}+\delta\sigma_{e2})]\, ,
\label{eq:X_prop}
\end{equation}
and
\begin{equation}
\left(\frac{\partial}{\partial t}+c\frac{\partial}{\partial
    z}\right) \delta ( Y_1^{\pi/2}+ Y_2^{\pi/2}) = -i g N [\delta\sigma_{1e}+\delta\sigma_{2e}-(\delta\sigma_{e1}+\delta\sigma_{e2})]\, .
\label{eq:P_prop}
\end{equation}
Thus, if Eq.~(\ref{eq:noise_is_zero}) is satisfied, then the right hand side of these
two equations becomes zero. This means that the sums of the fluctuations of the $\theta=0$
quadratures, $\delta(\hat Y_1^0+\hat Y_2^0)$, and of the
$\theta=\pi/2$ quadratures, $\delta(\hat Y_1^{\pi/2}+\hat Y_2^{\pi/2})$, of the fields do not change as they propagate.

The proposed procedure makes it possible to experimentally test for the violation of
Eqs.~(\ref{eq:vanloock_inequalities_no}) through the use of only field
measurements. Specifically, it requires the addition of the signals
from the two fields after propagation through the atomic medium to measure the total $\theta=0$
and $\theta=\pi/2$ quadrature noise spectra and verify that they are the same as  the ones of the input fields. If these conditions are
satisfied, then Eqs.~(\ref{eq:X_prop}) and (\ref{eq:P_prop}), with the
right hand side equal to zero, are satisfied, which implies that
Eq.~(\ref{eq:noise_is_zero}) is valid. This in turns means that it is possible to calculate $I_1(\omega)$, $I_2(\omega)$, and $I_3(\omega)$ through measurements of only field observables.

\section{Conclusion}
\label{sc:conclusions}
We show that EIT provides a novel mechanism for the deterministic generation of genuine tri-partite field-atom entanglement. Furthermore, the entanglement is generated through dissipation as the fields propagate through the EIT medium.  As a result, no state preparation of the atomic system is needed, which makes this approach very robust.  The tri-partite entanglement is generated when an initially entangled state of light, specifically bright twin beams, propagates through a medium composed of atoms in a $\Lambda$ configuration in the regime of EIT. We show that while in this regime the mean
values of the fields do not evolve once the dark state is established, their noise properties continue to evolve with propagation. The initial bi-partite entangled state of  light evolves into a state with genuine tri-partite entanglement between the two optical fields and the atomic ensemble. Finally, we identify an experimentally accessible set of measurements to verify the existence of the genuine tri-partite entanglement for the condition of equal input Rabi frequencies for the probe and control fields. In particular, the proposed scheme only requires measurements of the output quadrature noise spectra for each field and for combinations of the fields. The deterministic generation of hybrid atom-light entanglement will open up the door for the implementation of quantum networks and the distribution of entanglement between remote atomic-based memories and processors.

\section*{Appendix A: Full Expressions for Spectrum of Quadratures}\label{sc:app_a}

Through the use of the Heisenberg-Langevin equations and their conversion to $c$-number stochastic differential equations, we find that the most general analytical result for the noise spectrum of the individual field quadratures is given by
\begin{eqnarray}\label{eq:full_quad}
  {\mathcal S}_{22}^{\theta_1=0,\,
  \theta_2=0}(\omega,z)&=&1+\eta\left[-\frac{\Omega_2^2}{\Omega_1^2+\Omega_2^2}+\frac{\Omega_2^2 e^{2 r} (\Omega_1-\Omega_2)^2}{2 \left(\Omega_1^2+\Omega_2^2\right)^2}+\frac{\Omega_2^2 e^{-2 r} (\Omega_1+\Omega_2)^2}{2 \left(\Omega_1^2+\Omega_2^2\right)^2}\right.\nonumber\\&&\left.+e^{-Q_a z} \cos (Q_o z) \left(\frac{\Omega_1 \Omega_2 e^{-2 r}    \left(\Omega_1^2-\Omega_2^2\right)}{\left(\Omega_1^2+\Omega_2^2\right)^2}+\frac{e^{2 r} \left(\Omega_1 \Omega_2^3-\Omega_1^3 \Omega_2\right)}{\left(\Omega_1^2+\Omega_2^2\right)^2}\right)\right.\nonumber\\&&\left.+e^{-2 Q_a z} \left(-\frac{\Omega_1^2}{\Omega_1^2+\Omega_2^2}+\frac{\Omega_1^2 e^{-2 r} (\Omega_1-\Omega_2)^2}{2 \left(\Omega_1^2+\Omega_2^2\right)^2}+\frac{\Omega_1^2 e^{2 r} (\Omega_1+\Omega_2)^2}{2 \left(\Omega_1^2+\Omega_2^2\right)^2}\right)\right]
\end{eqnarray}
with
\begin{equation}
  Q_a= \frac{N g^2(\omega-\omega_a)^2\gamma/2}{c \left(\left(\Omega_1+\Omega_2\right)^2/4-\left(\omega-\omega_a\right)^2\right)^2+\left(\omega-\omega_a\right)^2\gamma^2/4}\, ,
    \label{eq:Qa_t}
\end{equation}
\begin{equation}
    Q_o= \frac{N g^2\left(\omega-\omega_a\right)\left(\Omega_1^2+\Omega_2^2-\left(\omega-\omega_a\right)^2\right)}{c \left(\left(\Omega_1+\Omega_2\right)^2/4-\left(\omega-\omega_a\right)^2\right)^2+\left(\omega-\omega_a\right)^2\gamma^2/4}\, , \label{eq:Qo}
  \end{equation}
  and where $\Omega_i=g\alpha_i$ and $c$ is the speed of light in vacuum. Corresponding analytical expressions for the cross-correlations of the field quadrates were also obtained, however these expressions are too complicated to present. In
  the derivation of these expressions we have  assumed that $g_1=g_2\equiv g$.
  The quadrature for
  ${\mathcal S}_{11}^{\theta_1=0,\, \theta_2=0}(\omega,z)$ is obtained
  by interchanging, in Eq.~(\ref{eq:full_quad}), labels $1$ and
  $2$. The quadrature
  ${\mathcal S}_{ii}^{\theta_1=\pi/2,\, \theta_2=\pi/2}(\omega,z)$ is
  obtained by changing the sign of $r$ in Eq.~(\ref{eq:full_quad}).

\section*{Funding}
This research was supported by Grant No. UNAM-DGAPA-PAPIIT
IG100518 and IG101421.


\begin{thebibliography}{10}
\newcommand{\enquote}[1]{``#1''}


\bibitem{rsta.2003.1227}
J.~P. Dowling and G.~J. Milburn, \enquote{Quantum technology: the second
  quantum revolution,} Philos. Trans. R. Soc. A
  \textbf{361}, 1655--1674 (2003).

\bibitem{PhysRevX.9.041042}
S.~Ecker, F.~Bouchard, L.~Bulla, F.~Brandt, O.~Kohout, F.~Steinlechner,
  R.~Fickler, M.~Malik, Y.~Guryanova, R.~Ursin, and M.~Huber,
  \enquote{Overcoming noise in entanglement distribution,}
  Phys. Rev. X \textbf{9}, 041042 (2019).

\bibitem{RN1173}
T.~D. Ladd, F.~Jelezko, R.~Laflamme, Y.~Nakamura, C.~Monroe, and J.~L.
  O’Brien, \enquote{Quantum computers,} Nature
  \textbf{464}, 45--53 (2010).

\bibitem{9149990}
D.~{Cuomo}, M.~{Caleffi}, and A.~S. {Cacciapuoti}, \enquote{Towards a
  distributed quantum computing ecosystem,} IET Quantum
  Commun. \textbf{1}, 3--8 (2020).

\bibitem{acsphotonics.9b00250}
B.~J. Lawrie, P.~D. Lett, A.~M. Marino, and R.~C. Pooser, \enquote{Quantum
  sensing with squeezed light,} ACS Photon.
  \textbf{6}, 1307--1318 (2019).

\bibitem{NatPhys.16.281}
X.~Guo, C.~R. Breum, J.~Borregaard, S.~Izumi, M.~V.~Larsen, T.~Gehring,
  M.~Christandl, J.~S. Neergaard-Nielsen, and U.~L. Andersen,
  \enquote{Distributed quantum sensing in a continuous-variable entangled
  network,} Nat. Phys. \textbf{16}, 281–284 (2020).

\bibitem{Zhuang_2020}
Q.~Zhuang, J.~Preskill, and L.~Jiang, \enquote{Distributed quantum sensing
  enhanced by continuous-variable error correction,} New
  J. Phys. \textbf{22}, 022001 (2020).

  \bibitem{McCormick:08}
C.~F. McCormick, A.~M. Marino, V. Boyer, and P.~D. Lett, \enquote{Strong low-frequency quantum correlations from a four-wave mixing amplifier,} Phys. Rev. A \textbf{78}, 043816 (2008).

\bibitem{Boyer_2008}
V.~Boyer, A.~M. Marino, R.~C. Pooser, and P.~D. Lett, \enquote{Entangled images
  from four-wave mixing,} Science \textbf{321},
  544--547 (2008).

\bibitem{Ding:15}
D.-S. Ding, W. Zhang, S. Shi, Z.-Y. Zhou, Y. Li, B.-S. Shi. and G.-C. Guo, \enquote{Hybrid-cascaded generation of tripartite telecom photons using an atomic ensemble and a nonlinear waveguide,} Optica
  \textbf{2}, 642--645 (2015).

\bibitem{Wang:17}
H. Wang, C. Fabre, and J. Jing, \enquote{Single-step fabrication of scalable multimode quantum resources using four-wave mixing with a spatially structured pump,} Phys. Rev. A
  \textbf{95}, 051802(R) (2017).

\bibitem{Park:19}
J. Park, H. Kim, and H.~S. Moon, \enquote{Polarization-entangled photons from a warm atomic ensemble using a Sagnac interferometer,} Phys. Rev. Lett. \textbf{122}, 143601 (2019).

\bibitem{Wang:20}
K. Wang, D.-S. Ding. W. Zhang, Q.-Y. He, G.-C. Guo, and B.-S. Shi, \enquote{Experimental demonstration of Einstein-Podolsky-Rosen entanglement in rotating coordinate space,} Sci. Bull.
  \textbf{65}, 280--285 (2020).

\bibitem{PhysRevLett.124.010510}
T.~van Leent, M.~Bock, R.~Garthoff, K.~Redeker, W.~Zhang, T.~Bauer,
  W.~Rosenfeld, C.~Becher, and H.~Weinfurter, \enquote{Long-distance
  distribution of atom-photon entanglement at telecom wavelength,}
  Phys. Rev. Lett. \textbf{124}, 010510 (2020).

\bibitem{RN1172}
C.~Simon, M.~Afzelius, J.~Appel, A.~Boyer de~la Giroday, S.~J. Dewhurst,
  N.~Gisin, C.~Y. Hu, F.~Jelezko, S.~Kröll, J.~H. Müller, J.~Nunn, E.~S.
  Polzik, J.~G. Rarity, H.~De~Riedmatten, W.~Rosenfeld, A.~J. Shields,
  N.~Sköld, R.~M. Stevenson, R.~Thew, I.~A. Walmsley, M.~C. Weber,
  H.~Weinfurter, J.~Wrachtrup, and R.~J. Young, \enquote{Quantum memories,}
  Eur. Phys. J. D \textbf{58}, 1--22 (2010).

\bibitem{09500340.2016.1148212}
K.~Heshami, D.~G. England, P.~C. Humphreys, P.~J. Bustard, V.~M. Acosta,
  J.~Nunn, and B.~J. Sussman, \enquote{Quantum memories: emerging applications
  and recent advances,} J. Mod. Opt.  \textbf{63}, 2005--2028 (2016).

\bibitem{RevModPhys.83.33}
N.~Sangouard, C.~Simon, H.~de~Riedmatten, and N.~Gisin, \enquote{Quantum
  repeaters based on atomic ensembles and linear optics,}
  Rev. Mod. Phys. \textbf{83}, 33--80 (2011).

\bibitem{Ren:12}
C. Ren and H.~F. Hofmann, \enquote{Clock synchronization using maximal multipartite entanglement,}
  Phys. Rev. A \textbf{88}, 014301   (2012).

\bibitem{Eldredge:18}
Z. Eldredge, M. Foss-Feig, J.~A. Gross, S.~L. Rolston, and A.~V. Gorshkov, \enquote{Optimal and secure measurement protocols for quantum sensor networks,}
  Phys. Rev. A \textbf{97}, 042337  (2018).

\bibitem{Hillery:99}
M. Hillery, V. Bužek, and A. Berthiaume, \enquote{Quantum secret sharing,}
  Phys. Rev. A \textbf{59}, 1829 (1999).

\bibitem{Zhu:15}
C.~Zhu, F.~Xu, and C.~Pei, \enquote{W-state analyzer and multi-party measurement-device-independent quantum key distribution,}
  Sci. Rep. \textbf{5}, 17449 (2015).

\bibitem{RevModPhys.77.633}
M.~Fleischhauer, A.~Imamoglu, and J.~P. Marangos, \enquote{Electromagnetically
  induced transparency: Optics in coherent media,} Rev.
  Mod. Phys. \textbf{77}, 633--673 (2005).

\bibitem{rv:marangos}
J.~P. Marangos, \enquote{Electromagnetically induced transparency,}
  J. Mod. Opt. \textbf{45}, 471--503 (1998).

\bibitem{rv:richter}
M.~Fleischhauer and T.~Richter, \enquote{Pulse matching and correlation of
  phase fluctuations in \ensuremath{\Lambda} systems,}
  Phys. Rev. A \textbf{51}, 2430--2442 (1995).

\bibitem{rv:pablomarc2}
P.~Barberis-Blostein and M.~Bienert, \enquote{Propagation of small fluctuations
  in electromagnetically induced transparency: Influence of doppler width,}
  Phys. Rev. A \textbf{79}, 063824 (2009).

\bibitem{PhysRevA.81.043802}
S.~Pielawa, L.~Davidovich, D.~Vitali, and G.~Morigi, \enquote{Engineering
  atomic quantum reservoirs for photons,} Phys. Rev. A
  \textbf{81}, 043802 (2010).

\bibitem{PhysRevLett.107.080503}
H.~Krauter, C.~A. Muschik, K.~Jensen, W.~Wasilewski, J.~M. Petersen, J.~I.
  Cirac, and E.~S. Polzik, \enquote{Entanglement generated by dissipation and
  steady state entanglement of two macroscopic objects,}
  Phys. Rev. Lett. \textbf{107}, 080503 (2011).

\bibitem{PhysRevA.83.052312}
C.~A. Muschik, E.~S. Polzik, and J.~I. Cirac, \enquote{Dissipatively driven
  entanglement of two macroscopic atomic ensembles,}
  Phys. Rev. A \textbf{83}, 052312 (2011).

\bibitem{Stannigel.2012}
K.~Stannigel, P.~Rabl, and P.~Zoller, \enquote{Driven-dissipative preparation
  of entangled states in cascaded quantum-optical networks,}
  New J. Phys. \textbf{14}, 063014 (2012).

\bibitem{lb:scully}
M.~Scully and M.~S. Zubairy, \emph{Quantum Optics} (Cambridge University Press,
  1997).

\bibitem{RevModPhys.81.865}
R.~Horodecki, P.~Horodecki, M.~Horodecki, and K.~Horodecki, \enquote{Quantum
  entanglement,} Rev. Mod. Phys. \textbf{81}, 865--942
  (2009).

\bibitem{van_loock_detecting_2003}
P.~van Loock and A.~Furusawa, \enquote{Detecting genuine multipartite
  continuous-variable entanglement,} Phys. Rev. A
  \textbf{67}, 052315 (2003).

\bibitem{PhysRevA.90.062337}
R.~Y. Teh and M.~D. Reid, \enquote{Criteria for genuine $n$-partite
  continuous-variable entanglement and einstein-podolsky-rosen steering,}
  Phys. Rev. A \textbf{90}, 062337 (2014).

\bibitem{PhysRevLett.97.140504}
A.~S. Villar, M.~Martinelli, C.~Fabre, and P.~Nussenzveig, \enquote{Direct
  production of tripartite pump-signal-idler entanglement in the
  above-threshold optical parametric oscillator,} Phys.
  Rev. Lett. \textbf{97}, 140504 (2006).

\bibitem{PhysRevLett.84.2722}
L.-M. Duan, G.~Giedke, J.~I. Cirac, and P.~Zoller, \enquote{Inseparability
  criterion for continuous variable systems,} Phys. Rev.
  Lett. \textbf{84}, 2722--2725 (2000).

\bibitem{rv:davidovich}
L.~Davidovich, \enquote{Sub-poissonian processes in quantum optics,}
  Rev. Mod. Phys. \textbf{68}, 127--173 (1996).

\bibitem{rv:pablomarc}
P.~Barberis-Blostein and M.~Bienert, \enquote{Opacity of electromagnetically
  induced transparency for quantum fluctuations,} Phys.
  Rev. Lett. \textbf{98}, 033602 (2007).

\bibitem{rv:dantan4}
A.~Dantan, A.~Bramati, and M.~Pinard, Phys. Rev. A
  \textbf{71}, 043801 (2005).

\bibitem{RevModPhys.81.1727}
M.~D. Reid, P.~D. Drummond, W.~P. Bowen, E.~G. Cavalcanti, P.~K. Lam, H.~A.
  Bachor, U.~L. Andersen, and G.~Leuchs, \enquote{Colloquium: The
  einstein-podolsky-rosen paradox: From concepts to applications,}
  Rev. Mod. Phys. \textbf{81}, 1727--1751 (2009).

\bibitem{Dowran_2018}
M.~Dowran, A.~Kumar, B.~J. Lawrie, R.~C. Pooser, and A.~M. Marino,
  \enquote{Quantum-enhanced plasmonic sensing,} Optica
  \textbf{5}, 628--633 (2018).

\end{thebibliography}
\end{document}